\documentstyle[12pt,epsf]{article}  
\newcommand{\Aa}{{\cal A}}

\newcommand{\Ss}{{\cal S}}

\newcommand{\wt}{\widetilde}
\newcommand{\tphi}{\wt\phi}
\newcommand{\tN}{\wt N}
\newcommand{\wh}{\widehat}

\newcommand{\hatn}{\wh n}

\newcommand{\be}{\begin{equation}}
\newcommand{\ee}{\end{equation}}
\newcommand{\ben}{\begin{eqnarray}\displaystyle}
\newcommand{\een}{\end{eqnarray}}
\newcommand{\refb}[1]{(\ref{#1})}

\newcommand{\sectiono}[1]{\section{#1}\setcounter{equation}{0}}

\begin{document}

{}~ \hfill\vbox{\hbox{hep-th/0102071}
\hbox{UUITP-01/01}}\break 

\vskip 3.0cm

\centerline{\large \bf  Mode Interactions of the Tachyon Condensate}
\vspace*{4.0ex}

\centerline{\large \bf in $p$-adic String Theory}
\vspace*{1.5ex}

\vspace*{8.0ex}

\centerline{\large \rm Joseph A. Minahan\footnote{E-mail: joseph.minahan@teorfys.uu.se}}
\vspace*{2.5ex}
\centerline{\large \it Department of Theoretical Physics}
\centerline{\large \it Box 803, SE-751 08 Uppsala, Sweden}
\vspace*{3.0ex}

\vspace*{2.5ex}

\vspace*{4.5ex}
\medskip
\centerline {\bf Abstract}

\bigskip
We study the fluctuation modes for lump solutions of the tachyon effective
potential in $p$-adic open string theory. We find a discrete spectrum
with equally spaced mass squared levels. 
We also find that the interactions derived
from this field theory are consistent with $p$-adic string amplitudes for
excited string states.
\vfill \eject
\baselineskip=17pt

\sectiono{Introduction}

Sen's conjectures \cite{senconj} concerning tachyon condensation and 
D-brane decay have been widely tested using cubic string field theory
\cite{cubiccalc} or the superstring analog \cite{superchecks}.
There have also been many promising attempts to find an exact
verification using cubic string field theory \cite{insight,0012251}.

Field theory models of tachyon dynamics are also useful in
understanding the qualitative picture of  tachyon condensation.
One such model was proposed in \cite{0008231}, which was constructed by
looking for lump solutions with 
a discrete fluctuation spectrum.  
This model is in some sense a generalization of a purely cubic 
model \cite{0008227} and was later shown to be the two derivative limit of
boundary string field theory \cite{9210065,0009103,0009148,0009191}.  
An analagous model
was constructed for the superstring \cite{0009246} and it too was later
shown to be the two derivative limit of an action derived from boundary
string field theory \cite{0010108}

The authors of \cite{0009103} pointed out that the model in \cite{0008231}
can be obtained as the  $p\to1$ limit of $p$-adic string theory.  
The p-adic string  describes tachyon dynamics in a vastly simpler context
than the ordinary bosonic string \cite{0003278}.  In particular,
Ghoshal and Sen showed that the open string dynamics disappear at the
stable vacuum.  They also showed that the lump solutions of the effective
field theory satisfy constant descent relations \cite{0003278}. 
One can ask how other ideas of tachyon condensation can be
tested using the $p$-adic string.

The $p$-adic string tree amplitudes are found by replacing the integrals over 
real numbers normally found in bosonic string theory, 
with integrals over $p$-adic numbers
\cite{FO,Volovich,Grossman,FrOk,BFOW,FrOk2}.  The resulting amplitudes
are much simpler than in bosonic string theory.  For example,
the $N$-point tachyon
amplitude only has tachyon poles.
Using these amplitudes, the authors of \cite{BFOW} were able to construct
the exact tree-level action for the tachyon field.  This action is
non-local, with a potential that is unbounded below, but with a local
minimum.  The equations of motion have
 non-trivial stationary classical solutions, where the tachyon field 
lies at the local minimum at spatial infinity.  These solutions
were later interpreted
to be the analogs of D-branes \cite{0003278}.

Since the tachyon amplitude only has tachyon poles, one might think that a
consistent $p$-adic string theory only needs the tachyon field.  In fact,
it is not clear that one can even put in many of the other fields
familiar from bosonic string theory.  For instance,  there has been no
successful attempt  to put in  gauge fields such that the resulting
amplitudes are gauge invariant.  Gauge invariance requires that the on-shell
amplitudes be zero when a polarization vector $\xi_\mu$ is replaced by the 
momentum vector
$k_\mu$.  In bosonic string theory, this works because
the vertex operator for the gauge boson,
\begin{equation}\label{vertop}
V=\zeta\cdot\partial_t X e^{ik\cdot X}
\end{equation}
where $\partial_t$ refers to the tangential derivative along the boundary of
the world-sheet, becomes a total derivative.  However,
for the $p$-adic numbers, 
there is no well defined notion of a derivative operator
\cite{BFOW}, hence there is no guarantee that the amplitude  is zero
after $p$-adic integration.  Indeed, if one considers the correlation functions
of  the vertex operators in \refb{vertop}, 
replacing the real number coordinates
on the boundary of the string world sheet with $p$-adic numbers,  then
after  $p$-adic integration one finds an amplitude that is not gauge
invariant.  

Instead of showing this explicitly, let us give another argument why
gauge fields are likely to be absent.
In \cite{BFOW} it was shown how
to insert Chan-Paton factors into the $p$-adic tachyon
 amplitudes.  
In bosonic string theory, when Chan-Paton factors are inserted, 
nonabelian gauge poles appear in the tree amplitudes, since the tachyons have
gauge indices.  One can also show that these amplitudes are consistent with
the tachyon-tachyon-gauge three point couplings.
But in the $p$-adic
amplitudes, even after inserting Chan-Paton factors, the only poles are
tachyon poles.  If a gauge vertex operator existed, then the three point
coupling would be the same as for the bosonic string, 
since there is no integration
over a number field.  This would then lead to an inconsistency with
the higher point amplitudes.  Therefore, for the $p$-adic string,
 the Chan-Paton symmetries are
global symmetries only. 

Nevertheless, the existence of lump solutions must lead to other $p$-adic
string states.  If one examines the effective field theory around the lump
solution, one finds fluctuation modes and the question is how do these
modes fit into the $p$-adic string picture.  The lowest such mode is a
tachyon \cite{FN,0003278}, while the next mode is the scalar zero-mode
\cite{0003278}.  

In this paper we will show that the remaining modes
lie in an infinite tower of states, with a discrete spectrum and evenly
spaced mass squared levels, for any value of $p$.  From the tachyon effective
action, one can compute the effective action for the fluctuation modes.
  From this effective action one can directly compute tree-level scattering
amplitudes.

In \cite{0011226} it was conjectured that in bosonic string theory 
the 
fluctuations of the tachyon field 
about a codimension $d$
 lump solution correspond to the open string
states
\begin{equation}\label{tachmodes}
\prod_{i=1}^d (\alpha_{-1}^i)^{m_i}|0\rangle,
\end{equation}
where the product is over the transverse directions to the brane.  The $m_i$
are nonnegative integers and $\alpha_{-1}^i$ is a string oscillator transverse
to the brane.  
One way to test the conjecture of \cite{0011226} is to compute the amplitudes
for the states in \refb{tachmodes} using
$p$-adic integration, and compare this to the amplitudes found from the 
$p$-adic
effective action.  We do this and find agreement between the two results.

In section 2 we study the fluctuation modes about the $p$-adic lump solutions
and find their spectrum and compute their interactions.  In section 3 
we compute the $p$-adic string amplitudes for the vertex operators in
\refb{vertopfm}.  In doing this we keep track of combinatoric
factors and simply borrow the results from 
\cite{FO,Volovich,Grossman,FrOk,BFOW,FrOk2} for tachyonic amplitudes, 
without encountering any new types of
$p$-adic integrals.  In section 4 we give a brief
discussion.

\sectiono{Mode interactions from the tachyon field theory}

The starting point for this analysis is the tachyon effective action
for the $p$-adic string \cite{BFOW}
\begin{equation}\label{effact}
\Ss=-\frac{1}{g^2}\frac{p^2}{p-1}\int d^Dx
\left[\frac{1}{2}\phi p^{-\frac{1}{2}\Box}\phi-
\frac{1}{p+1}\phi^{p+1}\right].
\end{equation}
While this action was derived for $p$ a prime number, it appears that $p$
can be continued to any positive real number.
The equation of motion for the $\phi$ field is easily derived from \refb{effact}
and is
\begin{equation}\label{eom}
p^{-\frac{1}{2}\Box}\phi=\phi^p.
\end{equation}
The perturbative open string vacuum is at $\phi=1$, while the solution at
the local minimum
$\phi=0$ has no open string fluctuations, since all poles are absent.
One  advantage of studying the $p$-adic string is that the strong
coupling problems that arise at the local minimum in boundary string field
theory \cite{0012081} are not a problem here \footnote{Nevertheless, one 
could argue that
the problem is alleviated by going to the weak string coupling limit.  If
the coupling is absorbed into the tachyon field $\phi$, then the potential
is $V(\phi)=-\frac{1}{4}\phi^2\ln(g^2\phi^2)$.  Hence the mass of the field
near the local minimum
is roughly $-\frac{1}{2}\ln(g^2\phi^2)$ while the coupling terms have no
$g$ dependence.  Hence, we can have a large mass and small couplings by 
keeping $\phi$ large, but $g\phi$ small.} so long as $p$ is an integer greater
than 1 \cite{0012251}.  

There are also nontrivial static solutions of \refb{eom}.  In particular,
there are codimension $d$ lump solutions \cite{BFOW,0003278}
\begin{equation}\label{lump}
\phi(x_i)=\prod_{i=1}^d f(x_i),
\end{equation}
where $f(x)$ is the gaussian
\begin{equation}\label{fdef}
f(x)\equiv p^{\frac{1}{2(p-1)}}
\exp\left(-\frac{1}{2}\frac{p-1}{p\ln p}x^2\right).
\end{equation}
This follows from the identity
\begin{equation}\label{iden1}
\Aa f(x)= (f(x))^p,\qquad\qquad\Aa\equiv p^{-\frac{1}{2}
\frac{\partial^2}{\partial x^2}}.
\end{equation}
Plugging the solution in \refb{lump} back into \refb{effact} one can compute
the tension of the lump solutions.  In particular, one finds that the ratio
of  tensions satisfies the descent relation \cite{0003278}
\begin{equation}\label{Tratio}
\frac{T_{D-d-1}}{T_{D-d-2}}=\left(\frac{2\pi p^{\frac{2p}{p-1}}\ln p}{p^2-1}
\right)^{-1/2}.
\end{equation}
Note that this ratio, as well as all tree amplitudes are invariant under 
$p\to 1/p$ \cite{FrOk2}.  One can show this by taking the equation of motion
in \refb{eom} and substituting $\phi=\chi^{1/p}$.  Then one finds the same
equation for $\chi$, but with $p$ replaced by $1/p$. 

Next consider the fluctuations about the lumps.  For what follows, we will 
assume that $d=1$, but the generalization for arbitrary $d$ is straightforward.
Hence, to quadratic order in the fluctuations, the action in \refb{effact}
becomes
\begin{equation}\label{flucact}
S_f=-\frac{1}{g^2}\frac{p^2}{p-1}\int d^{D-1}ydx
\frac{1}{2}\tphi(y,x)\left[ p^{-\frac{1}{2}\left(\Box_{\parallel}+
\frac{\partial^2}{\partial x^2}\right)}
-
p(f(x))^{p-1}
\right]\tphi(y,x),
\end{equation}
where $y$ are the coordinates along the lump world-brane, 
$\tphi(y,x)=\phi(y,x)-f(x)$,
and $\Box_\parallel$ is the d'Alembertian 
along the lump world-brane coordinates.
To find the eigenmodes, we use the following identity
\begin{equation}\label{iden2}
[\Aa,\ x]=-\ln p\frac{\partial}{\partial x}\ \Aa.
\end{equation}
Thus we find
\begin{equation}\label{aa}
\Aa\ x^n f(x)=p^n Q_n(x) (f(x))^p,
\end{equation}
where $Q_n(x)$ is a polynomial of the form
\begin{equation}\label{Qrel}
Q_n(x)=x^n+C_{n-2}x^{n-2} + C_{n-4}x^{n-4}+\dots.
\end{equation}
Hence, using \refb{aa}, \refb{Qrel} and the fact that $\Aa$ is a hermitian
operator, we see that
\begin{equation}\label{aa2}
\Aa\ H_n(\alpha x) f(x)=p^n H_n(\alpha x) (f(x))^p,
\end{equation}
 where 
\begin{equation}\label{aldef}
\alpha=\sqrt{\frac{1}{2}\frac{p^2-1}{p\ln p}},
\end{equation}
and where $H_n(\xi)$ are the Hermite polynomials with the usual normalization
\begin{equation}\label{Hnorm}
\int_{-\infty}^{+\infty}d\xi \exp\left(-\xi^2\right)
H_n(\xi)H_m(\xi)=
\pi^{\frac{1}{2}}2^nn! \delta_{nm}.
\end{equation}

We now write the $\tphi$ fields in the form
\begin{equation}\label{tphieq}
\tphi(y,x)=\sum_{n=0}^\infty \psi_n(y)2^{-\frac{n}{2}}H_n(\alpha x)
f(x),
\end{equation}
therefore, \refb{flucact} becomes
\begin{equation}\label{newfluc}
S_f=-\frac{1}{g^2}\frac{p^2}{p-1}
\left[\frac{2\pi p\ln p}{p^2-1}\right]^{\frac{1}{2}}\int d^{D-1}y
\sum_n \frac{n!}{2}
\psi_n(y)\left[ p^{\left(n-\frac{1}{2}\Box_{\parallel}\right)}-
p\right]\psi_n(y).
\end{equation}
Hence, the fluctuation modes $\psi_n(x)$ have masses squared given by
\begin{equation}
m^2=2(n-1).
\end{equation}
As in the normal bosonic string, the spectrum is discrete with equally spaced
levels.
The lowest such mode has $n=0$ and is a tachyon, as was first discussed in 
\cite{FN}.  Its mass squared is the same as the effective tachyon mass in
\refb{effact}.  The
next highest mode is at $n=1$, and is the massless mode discussed in
\cite{0003278}.  

Let us next consider the interaction terms.  To this end 
we make the substitution
\begin{equation}\label{phiexp}
\phi(y,x)=f(x)+\sum_n\psi_n(y)2^{-\frac{n}{2}}H_n(\alpha x)f(x)
\end{equation}
into the interaction term in \refb{effact}.
Therefore, the interacting part of the lump action is
\begin{equation}\label{Sint}
S_{int}=\frac{1}{g^2}\frac{p^2}{p-1}
\left[\frac{2\pi p\ln p}{p^2-1}\right]^{\frac{1}{2}}\int d^{D-1}y
\sum_{\ell=3}^{p+1}\frac{p\ !}{\ell\ !(p+1-\ell)!}\sum_{n_1}
\cdots\sum_{n_\ell} 
A_{n_1n_2\dots n_\ell}\prod_{i=1}^\ell\psi_{n_i}(y),
\end{equation}
where
\begin{equation}\label{Aeq}
A_{n_1n_2\dots n_\ell}=\left[\frac{2\pi p\ln p}{p^2-1}\right]^{-\frac{1}{2}}
\int_{-\infty}^{+\infty}dx \exp\left(-\frac{1}{2}\frac{p^2-1}{p\ln p}x^2\right)
\prod_{i=1}^\ell 2^{-\frac{n_i}{2}}H_{n_i}(\alpha x).
\end{equation}

To evaluate the expression in \refb{Aeq}, we note that $H_n(\alpha x)$ 
can be written in operator form

\begin{equation}\label{Hop}
2^{-\frac{n}{2}}H_n(\alpha x)=\ :(a+a^\dag)^n:
\end{equation}
where
\begin{equation}\label{adef}
a=\alpha x +\frac{1}{2\alpha}\frac{\partial}{\partial x}.
\end{equation}
Therefore
\begin{equation}
\qquad\qquad [a,a^\dag]=1
\end{equation}
and the equation in \refb{Aeq} becomes\footnote{
The form for $H_n(\alpha x)$ in \refb{Hop}  follows from the fact that
$:(a+a^\dagger)^n:$ commutes with $x=\frac{1}{2\alpha}(a+a^\dagger)$ and
that $:(a+a^\dagger)^n:|0\rangle=(a^\dagger)^n|0\rangle$.}
\begin{equation}\label{Aeq2}
A_{n_1n_2\dots n_\ell}=\langle0|\prod_i^\ell :(a+a^\dag)^{n_i}:|0\rangle,
\end{equation}
where the oscillator vacuum state is normalized to be 
\begin{equation}
\langle0|0\rangle=1.
\end{equation}

Therefore, the coefficient $A_{n_1n_2\dots n_\ell}$ essentially counts
the number of possible ways to contract the $\ell$ fields together.  To see
this more explicitly, let $N_{ij}$ be the number of contractions between
$:(a+a^\dag)^{n_i}:$ and $:(a+a^\dag)^{n_j}:$ in \refb{Aeq2}.  Since the 
operators are normal ordered $N_{ii}=0$.  The $N_{ij}$
satisfy the $\ell$ constraint conditions
\begin{equation}\label{Nijc}
n_i=\sum_{j\ne i} N_{ij}
\end{equation}
and for fixed $N_{ij}$, the total number of different ways to have this set
of contractions is
\begin{equation}\label{concount}
\frac{\prod_i n_i\ !}{\prod_{i<j} N_{ij}\ !}.
\end{equation}
Hence, the coefficients can be expressed as
\begin{equation}\label{Aeq3}
A_{n_1n_2\dots n_\ell}=
\sum_{N_{ij}}\frac{\prod_i n_i\ !}{\prod_{i<j} N_{ij}\ !},
\end{equation}
where the sum over all $N_{ij}$ is subject to the constraints in \refb{Nijc}.

\sectiono{$p$-adic amplitudes for excited modes}

In this section we consider the $N$-point amplitudes for excited $p$-adic
string modes.  There are certain points to keep in mind when constructing
these amplitudes.  First, the number of excited modes that could correspond
to fluctuation modes for a tachyon field is much smaller than the 
number of string modes in ordinary bosonic string theory.  This suggests two
possibilities.  Either an infinite number of fields need to be put in by hand
for the full effective field theory, 
or the number of string modes for $p$-adic string theory is vastly reduced
from ordinary bosonic string theory.  
The second possibility seems more likely,
given the problems previously discussed for gauge fields.

Nevertheless, 
while massless gauge fields and many other fields
might not exist, scattering amplitudes
for certain
string modes with polarizations transverse to the brane do appear to be
consistent.  For one thing, these are not gauge fields, so there is no
gauge invariance restriction.  For a second thing, the existence of tachyon 
lumps, which follows from the effective action derived from tachyon scattering
amplitudes seems to require the existence of such modes.

In \cite{0011226} it was suggested that for the two derivative truncation of 
boundary string field theory,
the tachyon field condensed to a codimension $d$ brane has fluctuations that
correspond to the open string states in \refb{tachmodes}.
The vertex operator for such a mode is
\begin{equation}\label{vertopfm}
V=\prod_{i=1}^d (\partial_n X^i)^{m_i} e^{ik\cdot X},
\end{equation}
where $\partial_n$ refers to the normal derivative from the boundary of the
string worldsheet.  It is important to note that $k_\mu$ points along the
brane coordinates, 
therefore correlation functions between vertex operators of this
sort do not have contributions from the contractions of $\partial_n X^i$
with $e^{ik\cdot X}$ terms.

Let us now specialize to the case where $d=1$ and let us consider the $N$-point
correlator
\begin{equation}\label{corr}
\langle \prod_i^N(\partial X(x_i))^{n_i}e^{ik_i\cdot X(x_i)}\rangle.
\end{equation}
Using the fact that 
\begin{equation}\label{Xcorr}
\langle X(z_i)X(z_j)\rangle=-\log|z_i-z_j|
\end{equation}
we find that the correlator in \refb{corr} is
\begin{equation}\label{corr2}
\sum_{N_{ij}} \prod_i^N n_i!\prod_{i<j} \frac{1}{N_{ij}!}
|x_{ij}|^{k_i\cdot k_j-2N_{ij}},
\end{equation}
where $N_{ij}$ counts the number of contractions of $\partial X(x_i)$ with
$\partial X(x_j)$.  As in the previous section, the $N_{ij}$ are subject
to a constraint
\begin{equation}\label{Nijc2}
n_i=\sum_{j\ne i} N_{ij}.
\end{equation}

Therefore, the $N$-point amplitude for these excited states is a sum over
tachyon like amplitudes, where each term in the sum has the combinatoric 
prefactor in \refb{concount} 
and where $k_i\cdot k_j$ is replaced with $k_i\cdot k_j-2N_{ij}$.  We
can then use information about $p$-adic tachyon amplitudes to find
the $p$-adic excited state amplitudes. 

The $p$-adic amplitudes are made up of a constant term independent of
the particles momenta and a sum over terms with poles \cite{BFOW}.
The constant term directly corresponds to the $N$-point interaction term
in \refb{Sint}.  Since, it is independent of $k_i\cdot k_j$, the constant
term is unchanged if $k_i\cdot k_j$ is replaced by $k_i\cdot k_j -2N_{ij}$.
Thus each term
in the sum in \refb{corr2} has the same constant term.  Hence, the general
$N$-point interaction term has an extra factor of
\begin{equation}\label{exfactor}
\sum_{N_{ij}} \frac{\prod_i^N n_{i}!}{\prod_{i<j}N_{ij}!}=A_{n_1n_2\dots n_N},
\end{equation}
precisely matching the result of the field theory analysis.

To complete the proof, we need to show that
the poles in the amplitude derived from the correlator in \refb{corr}
factorize  consistently with
the action in \refb{newfluc} and \refb{Sint}.  We proceed using an induction
argument.  If we isolate the amplitude to a particular pole term, then the
propagator divides the Feynman diagram into two parts (see figure 1).  

\begin{figure}[!ht]
\leavevmode
\begin{center}
\epsfbox{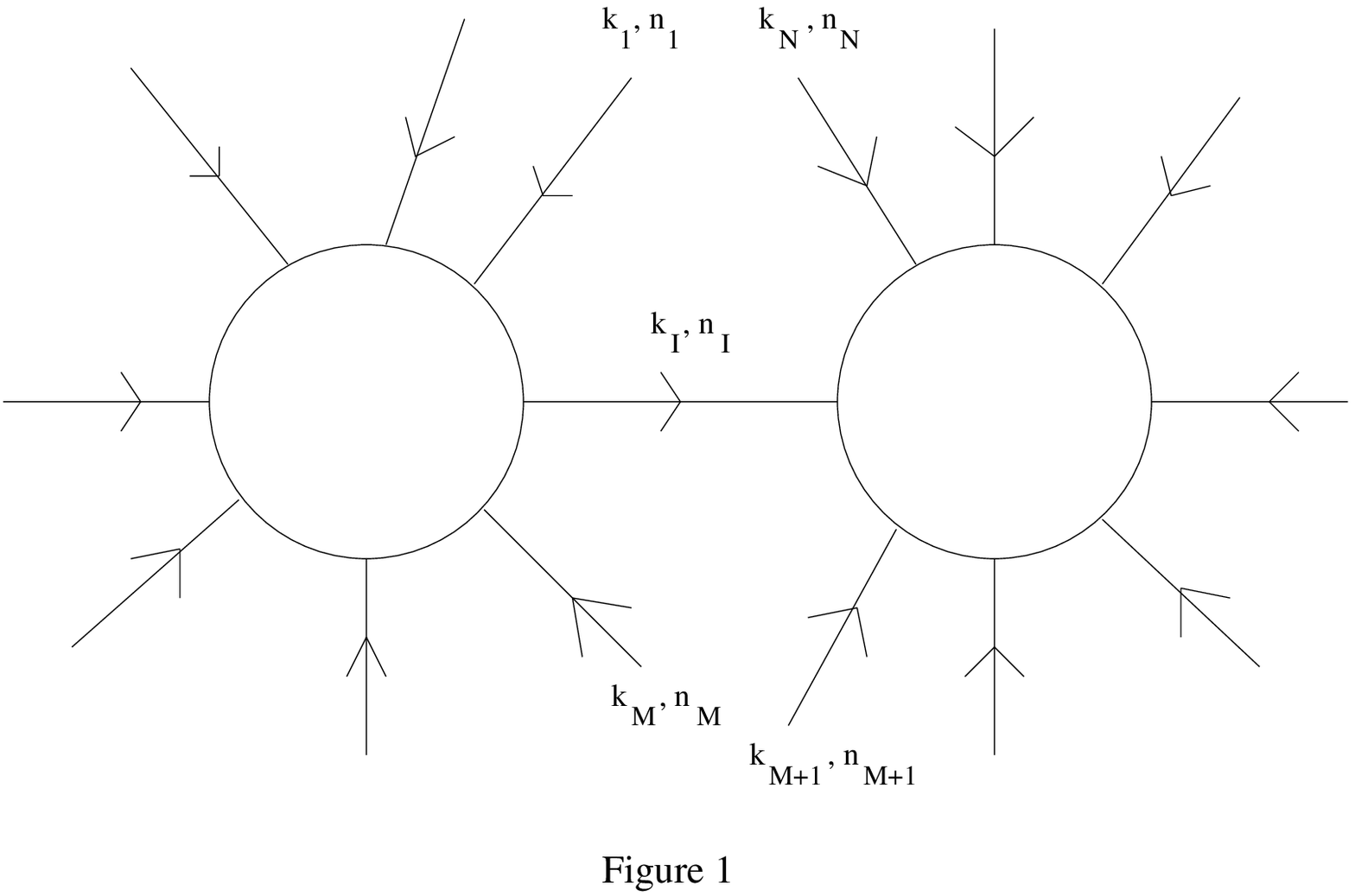}
\end{center}
\caption[]{\small Factorization of the amplitude with incoming momenta $k_i$
and mode number $n_i$ about a pole with
momentum $k_I$ and mode number $n_I$.}  \label{f1}
\end{figure}

Let us assume
that there are $M$ vertices on one side of the propagator
 and $L=N-M$ vertices on the
other side.  By assumption, the part of the amplitude containing this pole 
factorizes to the
form
\begin{equation}\label{factorize}
A(n_1,k_1;n_2,k_2;\dots n_M,k_M;n_I,k_I)\frac{1}{n_I!}
\frac{1}{p^{n_I+k_I^2/2}-p}
A(n_I,k_I;n_{M+1},k_{M+1}\dots n_N,k_N),
\end{equation}
where $A(n_1,k_1;n_2,k_2;\dots n_M,k_M;n_I,k_I)$ and 
$A(n_I,k_I;n_{M+1},k_{M+1}\dots n_N,k_N)$ refer to the lower point amplitudes
derived from \refb{newfluc} and \refb{Sint}
and $n_I$ and $k_I$ refer to the mode number and momentum
of the internal state.
Hence the momentum going through the propagator is
\begin{equation}\label{polemom}
k_I=\sum_i^{M}k_i=-\sum_{i=M+1}^N k_i,
\end{equation}
and so
\begin{eqnarray}
k_I^2&=&\sum_i^M k_i^2 + 2\sum_{i<j}^M k_i\cdot k_j=2\sum_i^M (1-n_i)
+ 2\sum_{i<j}^M k_i\cdot k_j\nonumber\\
&=&2M + 2\sum_{i<j}^M(k_i\cdot k_j-2M_{ij})-2\sum_i^M\sum_j^{L}\tN_{ij}
\end{eqnarray}
where 
\begin{eqnarray}
M_{ij}&=&N_{ij}\qquad\qquad i,j\le M\nonumber\\
\tN_{ij}&=&N_{i,j+M}\qquad\qquad i\le M, \ \ j\le L.
\end{eqnarray}
Comparing to a pure
tachyon amplitude, which only has tachyon poles, it is clear that the mass
squared of the pole is given by
\begin{equation}\label{polemass}
m_I^2=-2+\sum_{i}^M\sum_j^{L}\tN_{ij},
\end{equation}
and so the mode number for the pole is
\begin{equation}\label{intmode}
n_I=\sum_{i}^M\sum_j^{L}\tN_{ij}.
\end{equation}
By momentum conservation, we also have that 
\begin{equation}
k_I^2=2L + 2\sum_{i<j}^L(k_i\cdot k_j-2L_{ij})+
\sum_i^M\sum_j^{L}\tN_{ij},
\end{equation}
and hence we have the relation that
\begin{equation}
M + \sum_{i<j}^M(k_i\cdot k_j-2M_{ij})=
L + \sum_{i<j}^L(k_i\cdot k_j-2L_{ij}).
\end{equation}

There are $N(N-1)/2$ separate $N_{ij}$ terms.  However, the constraint in
\refb{Nijc2} reduces the number of independent terms to $N(N-3)/2$.  Since
we are isolating the amplitude to a particular pole, 
$n_I$ is fixed, reducing the number of independent terms by 1.
Finally, fixing all $M_{ij}$ and $L_{ij}$ reduces
the number of independent $N_{ij}$ terms by $(M+1)M/2-(M+1)$ and
$(L+1)L/2-(L+1)$, leaving  $(M-1)(L-1)$ independent terms.  These
independent terms are the $\tN_{ij}$, subject to the constraints
\begin{equation}\label{intrel}
M_{iI}\equiv n_i-\sum_{k\ne i}M_{ik}=\sum_k\tN_{ik}\qquad\qquad 
L_{jI}\equiv \hatn_j-\sum_{k\ne j}L_{jk}=\sum_k\tN_{kj}, 
\end{equation}
where $\hatn_i=n_{i+M}$.  The terms $M_{iI}$ and $L_{iI}$ count the
contractions of the external states with the internal state, and impose
$M+L-1$ constraints.

We now write the combinatoric factor in \refb{concount} as
\begin{equation}\label{facsplit}
\frac{\prod_i n_i!}{\prod_{i<j} N_{ij}!}
=\frac{\prod_i^M n_i!\prod_i^L \hatn_i!}
{\prod_{i<j}^M M_{ij}!\prod_{i<j}^L L_{ij}!\prod_{i,j} \tN_{ij}! }.
\end{equation}
Since $\tN_{ij}$ enters into the factorized amplitudes  only through the
$M_{iI}$ and $L_{jI}$ terms,
the complete factorized amplitude should consist of a sum
over all $(M-1)(L-1)$ independent $\tN_{ij}$.  It is convenient to use  as
the independent terms $\tN_{ij}$ with $i<M$ and $j<L$.  
Then the dependent terms
can be written as
\begin{eqnarray}\label{dependent}
\tN_{iL}&=&n_i-\sum_k^{L-1}\tN_{ik}-\sum_{k}M_{ik}\nonumber\\
\tN_{Mj}&=&\hatn_j-\sum_k^{M-1}\tN_{kj}-\sum_{k}L_{kj}\\
\tN_{ML}&=&n_M-\sum_k^{L-1}\tN_{Mk}-\sum_{k}M_{Mk}\nonumber\\
	&=&n_M-\sum_k^{L-1}\hatn_k+\sum_m^{M-1}\sum_k^{L-1}\tN_{mk}
-\sum_{k}M_{Mk}+\sum_{jk}L_{jk}.\nonumber
\end{eqnarray}

Thus, using this way to choose the dependent terms, we see that the factor
in \refb{facsplit} dependent on $\tN_{ij}$  is
\begin{equation}\label{Nijfac}
\frac{1}{\tN_{ij}!\tN_{Mj}!\tN_{iL}!\tN_{ML}!}.
\end{equation}
Starting with the sum over $\tN_{11}$, we have
\begin{eqnarray}\label{N11fac}
&&\sum_{\tN_{11}}\frac{1}{\tN_{11}!\tN_{M1}!\tN_{1L}!\tN_{ML}!}=\nonumber\\
&&\qquad\qquad 
\sum_{\tN_{11}}\left({\tN_{11}+\tN_{1L}\atop \tN_{11}}\right)
\left({\tN_{M1}+\tN_{ML}}\atop\tN_{1L}\right)\frac{1}{(\tN_{11}+\tN_{1L})!
(\tN_{M1}+\tN_{ML})!}.
\end{eqnarray}
Note that $(\tN_{11}+\tN_{1L})$ and $(\tN_{M1}+\tN_{ML})$ on the 
rhs of \refb{N11fac} are independent of
$\tN_{11}$. 
Using the identity
\begin{equation}\label{combiden}
\sum_m \left({N_1 \atop m}\right)\left({N_2\atop N_3-m}\right)=\left({N_1+N_2
\atop N_3}\right),
\end{equation}
we can write \refb{N11fac} as
\begin{equation}\label{N11sum}
\frac{(\tN_{11}+\tN_{1L}+\tN_{M1}+\tN_{ML})!}{
(\tN_{11}+\tN_{M1})!(\tN_{1L}+\tN_{ML})!
(\tN_{11}+\tN_{1L})!(\tN_{M1}+\tN_{ML})!}.
\end{equation}
The last two terms in the denominator of this expression have $\tN_{12}$
dependence.  Hence, after 
dividing \refb{N11sum} by $\tN_{12}!\tN_{M2}!$ and using \refb{combiden},
the sum over $\tN_{12}$ gives
\begin{equation}\label{N12sum}
\frac{(\tN_{11}+\tN_{12}+\tN_{1L}+\tN_{M1}+\tN_{M2}+\tN_{ML})!}{
(\tN_{11}+\tN_{M1})!(\tN_{12}+\tN_{M2})!(\tN_{1L}+\tN_{ML})!
(\tN_{11}+\tN_{12}+\tN_{1L})!(\tN_{M1}+\tN_{M2}+\tN_{ML})!}.
\end{equation}
Following this same procedure for the sums of all $\tN_{1j}$, $j<M$, results
in the factor
\begin{equation}
\frac{\left(\sum_j^L(\tN_{1j}+\tN_{Mj})\right)!}
{ \left(\sum_j^L\tN_{1j}\right)!\left(\sum_j^L\tN_{Mj}\right)!
\prod_j^L(\tN_{1j}+\tN_{Mj})!}
\end{equation}

One can now continue in a similar fashion for all  sums, where one uses the
identity
\refb{combiden} for each sum.  At the end, one finds that
\begin{equation}\label{Nijsum}
\sum_{N_{ij}}\prod_{i}^M\prod_j^L\frac{1}{\tN_{ij}!}=
\frac{\left(\sum_{ij}\tN_{ij}\right)!}{\prod_i\left(\sum_j\tN_{ij}\right)!
\prod_j\left(\sum_i\tN_{ij}\right)!}=
\frac {n_I!}{\prod_i M_{iI}!\prod_j L_{jI}!},
\end{equation}
where we used \refb{intmode} and \refb{intrel} for the last equality.
Thus, the amplitudes\hfill\break
 $A(n_1,k_1;n_2,k_2;\dots n_M,k_M;n_I,k_I)$ and
$A(n_I,k_I;n_{M+1},k_{M+1}\dots n_N,k_N)$ have combinatoric factors
\begin{equation}\label{A1}
\frac{\prod_i^M n_i!}{\prod_{i<j}^M M_{ij}!}\frac{n_I!}{\prod_i^M M_{iI}!}
\end{equation}
and
\begin{equation}\label{A2}
\frac{\prod_i^L \hatn_i!}{\prod_{i<j}^L L_{ij}!}\frac{n_I!}{\prod_i^L L_{iI}!}
\end{equation}
respectively.
Therefore, each part of the factorized amplitude has the same form for its
combinatoric factors as the original amplitude.  We can then continue 
factorizing poles until we are left with pole free amplitudes.  These amplitudes then have the combinatoric factors in \refb{exfactor}.  This completes the
proof.

\sectiono{Discussion}

We have shown that the effective field theory for the tachyon fluctuations
about a $p$-adic lump is consistent with tree-level 
$p$-adic string amplitudes.  This analysis is further evidence for the 
identification of tachyon fluctuations with specific open string states.
The analysis in \cite{0011226} went further, conjecturing the identification
of gauge fluctuations with other open string states.  The apparent absence of
gauge fields prevents us from testing this in the $p$-adic case.

There also have been attempts to generalize the closed string by replacing
integrals over the complex plane with integrals over $p$-adics, or more
precisely, with integrals over extensions of $p$-adic numbers 
\cite{FO,FW,Volovich,Grossman,FN2}.  The resulting amplitudes are almost
identical to the open string amplitudes, and in fact an effective field theory
was found for the tachyon, which is identical to the action in \refb{effact},
except that the d'Alembertian has a factor of $1/4$ in front of it instead
of $1/2$ \cite{FN2}.  But this  means that this action has lump solutions
with constant descent relations.   If this is really a closed string theory,
then it seems difficult to interpret the lumps as analogs of  D-branes.  
A possible 
explanation is that this construction is  the analog of the 
open string for {\it extended} $p$-adic fields.

If this is the case, then it begs the question of whether there is a $p$-adic
closed string.  The apparent absence of gauge fields on the $p$-adic
worldsheet might provide a clue.  It has been argued that in ordinary
string theory, the  closed strings
are flux tubes coming from these gauge fields 
\cite{9901159,0002223,0005031,0009061,0010240,0012081,0011009}.  
But if the gauge
fields are absent, then there might not be any closed strings to be concerned
about.

\bigskip
\noindent {\bf Acknowledgments}:
I thank Barton Zwiebach for many helpful discussions and for critically
reading the manuscript.  I also thank  the CTP at MIT for 
hospitality during the course of this work.
This work was
supported in part by the NFR.

\end{document}